\documentclass[a4paper, twocolumn]{article}
\usepackage{graphicx}
\usepackage{courier}
\usepackage{flushend,cuted}
\usepackage{amsmath}
\usepackage{url}
\usepackage{soul}
\usepackage{subcaption}
\usepackage{listings}

\title{Data Exfiltration via Multipurpose RFID Cards and Countermeasures}
\author{Zheng Zhou, Weiming Zhang, Nenghai Yu\\
zhou7905@mail.ustc.edu.cn; \{zhangwm, ynh\}@ustc.edu.cn;\\
University of Science and Technology of China\\
Key Laboratory of Electromagnetic Space Information\\ 
of the Chinese Academy of Sciences} 
\begin{document}
\maketitle
\begin{strip}
\begin{abstract}
Radio-frequency identification(RFID) cards are widely applied in daily human life. The information security of RFID cards, such as data confidentiality, tag anonymity, mutual authentication etc, has been fully studied.
In the paper, using the RFID cards in MIFARE Classic and DESFire families, a bidirectional covert channel via multipurpose RFID cards between service providers is built to leak sensitive data between two simulation systems. 
Furthermore, by calculations and experiments, the daily channel capacity to leak data of the channel is obtained. 
Although the storage capacity of a single RFID card is very small, a large user base can still bring about a considerable amount to leak data.
Then, the reasons for the existence of such channels are discussed. 
To eliminate this type of covert channels, a new authentication protocol between RFID cards and card readers are proposed.
Our experimental results show a significant security improvement in prevention of such covert communications while keeping user convenience.
\end{abstract}

\begin{flushleft}
\end{flushleft}
\end{strip}
\raggedend
%\flushend
\section{Introduction}

%RFID
Radio-frequency identification(RFID) technology was introduced in 1946.\cite{Herzog:2008:HEL:1205341} In the past 30 years, RFID have been developed rapidly, and play an important role in the automatic identification technology due to its advantages such as low cost, small size, quick identification and even battery-free. It also has a great economy effect. ``in 2017, the total RFID market will be worth \$11.2 billion, up from \$10.52 billion in 2016 and \$9.95 billion in 2015. This includes tags, readers and software/services for RFID labels, cards, fobs and all other form factors, for both passive and active RFID. IDTechEx forecast that to rise to \$14.9 billion in 2022.''\cite{RFIDForecasts}

%Galaxy
As a kind of product form, RFID cards are widely used in daily human life to keep us convenience. The main applications involve entrance guards, consumer stored-value cards and public traffic cards etc. 

%Star
However, the more RFID services are provided, the more RFID cards are needed. It is inconvenient if a user has to take so many RFID cards. Therefore, in a certain region such as a campus or a factory, an ``all-in-one'' technique is utilized to realize multiple RFID services by using one single card.
The main advantage of the ``all-in-one'' technique is that the count of cards is reduced, hence the cost is decreased. Meanwhile, users' cost of carry is reduced. However, there is a potential security risk that such a multi-purpose card can also be exploited as the payload of a covert channel.

%Universe
%Covert Channel
A \textit{covert channel} is a well-known way to transmit messages by circumventing security mechanisms. The definition of a covert channel was given by Lampson in 1973 to describe the leakage of data due to abuse of shared resources by processes with different privilege levels\cite{Lampson:1973:NCP:362375.362389}. In the last 40 years, the idea of a covert channel has been extended from a single host to a network, even to several networks that are physically separated from each other. The most famous single-host covert channel is exploited by Meltdown\cite{Meltdown} and Spectre\cite{Spectre} that cause a great security issue early in 2018. The network covert channels through which sensitive data are hidden in network packages with steganography were surveyed by Zander\cite{Zander2007-4317620} in 2007. A physically separated covert channel, also called an \textit{air-gapped} covert channel, leaks data directly via a physical emitting source such as an LED\cite{Loughry:2002:ILO:545186.545189, Sepetnitsky-2014-6975588, Guri2017LED, guri2017xled, Zhou8369122, ZHOU2019307}, an acoustic source\cite{hanspach2014covert, Malley-o2014bridging, lee2015various, Guri-Fansmitter-2016arXiv160605915G, Guri2017DiskFiltration} or an antenna\cite{kuhn1998soft, guri2014airhopper, guri2015gsmem, guri2016usbee, Matyunin2016Covert} etc. No any normal channel can be utilized for such air-gapped covert channels. Therefore, an air-gapped covert channel must be built by finding a new communication method.

%Limitation
Most researches on RFID security focus on lightweight algorithm, anonymity, integrity, confidentiality and efficiency. The situation under which two or more service providers collude with each other is ignored. 
If the service providers are in different levels, they can communicate with each others illegally by using the memory in the RFID cards together.
This type of malicious covert communications hide themselves into a normal business process. And the covert data transmitted illegally are stored in encrypted storage spaces in the card. Therefore, few method can detect them effectively.

%My Work
In the paper, a prototype on this type of covert channels between service providers is introduced. The effectiveness to leak data is evaluated by calculating the daily channel capacity. The reasons for the existence of such covert channels are discussed. A novel authentication protocol is proposed to prevent of such covert communications.

%Advantage
The new authentication protocol can be utilized to prevent the collusion of multiple service providers effectively. The risk of such covert channels is eliminated. Thus the security of RFID cards in the ``all-in-one'' mode is maintained.

%Stage V:
%Value
The new authentication protocol can also be referred to design a new generation of multi-purpose RFID cards used in the ``all-in-one'' mode. Meanwhile, the design idea of the new protocol can be extended to other similar application models.

The rest of the paper is organized as follows: Related works are described in Section \ref{RelativeWorks}. The technical background is described in Section \ref{BackgroundTechnology}. A prototype is introduced in Section \ref{AttackModel}. Section \ref{ResultandDiscussion} presents and discuss our experimental results. A new protocol against the attack is proposed in Section \ref{ProposedProtocol}. Finally, our conclusions are drawn in Section \ref{Conclusions}.

\section{Related Works}\label{RelativeWorks}
In the past 20 years, the study on RFID security focus on lightweight algorithms for the fast speed of hash functions. In 2005, Yang et al.\cite{yang2005mutual} proposed a mutual authentication protocol in a hostile surrounding. Not surprisingly, the protocol was compromised by replay attack. Luo et al.\cite{luo2005lightweight} introduced another mutual authentication protocol in the same year to extend the protocol presented by Ohkubo et al.\cite{ohkubo2003cryptographic} in using the two hash functions for updating secret values. The protocol was defeated by DoS and replay attacks. In 2006, Lee et al.\cite{lee2006rfid} proposed a mutual authentication scheme on synchronized secret information using both XOR(exclusive or) and hash chains to authenticate tags and readers. The scheme failed in a tag impersonation. In 2007, Kang and Lee\cite{kang2007study} introduced another mutual authentication protocol reducing the key length and communications. The protocol is also failed by a tag impersonation. Han et al.\cite{han2007mutual} proposed a mutual authentication protocol based on synchronized secret information. The protocol is vulnerable in replay attacks. Ha et al. presented LRMAP\cite{ha2007lrmap}, a resynchronous mutual authentication protocol that can recover synchronization between the database and the tag. The protocol can be compromised by a DoS attack. In 2008, Cai Qingling et al.\cite{qingling2008minimalist} introduced a minimalist mutual authentication protocol using CRC(Cyclic Redundancy Check). The protocol failed in a tag and reader impersonation. Tan et al.\cite{tan2008secure} proposed an authentication protocol without need for a central database. The protocol was also compromised by replay attack. Song et al.\cite{song2008rfid} introduced another authentication protocol. In 2009, Cai et al.\cite{cai2009attacks} studied several mutual authentication protocols and proposed revised protocols to eliminate the vulnerabilities. The revised protocols were also vulnerable by DoS and tag impersonations. In 2011, Piramuthu\cite{piramuthu2011rfid} surveyed 10 mutual authentication protocols from 2005 to 2009. In 2015, Cho et al.\cite{cho2015consideration} proposed a hash-based tag mutual authentication protocol. In 2018, Gope et al.\cite{gope2018lightweight}introduced an authentication scheme for distributed IoT infrastructure.

%RFID Card
In 2007, Heydt-Benjamin et al.\cite{Benjamin10.1007/978-3-540-77366-5_2} found the vulnerabilities in the first-generation
RFID-enabled credit cards.
Garcia\cite{Garcia5207633}showed that MIFARE Classic, a widely used smart card, was vulnerable to reverse engineering and man-in-the-middle attack in 2009.
D\"{u}zenli\cite{DUZENLI2015} proposed a novel approach that applies neural network forecasting to security for closed-loop prepaid cards based on low-cost technologies such as RFID and 1-Wire in 2015.
In 2017, Hossain et al.\cite{Hossain10.1007/978-3-319-72395-2_55} proposed a security solution of an RFID card through cryptography.
Xu et al.\cite{Xu2018} proposed a side-channel attack on a protected RFID card in 2018.

\section{Technical Background}\label{BackgroundTechnology}
\subsection{RFID Card}\label{RadioFrequencyCard}
RFID uses carrier waves of various bands, such as 120$\sim$150 kHz, 13.56 MHz, 433 MHz, 860$\sim$960 MHz, 2.45 GHz and 5.8 GHz. 
Among them, the band on 120$\sim$150 kHz is used for animal identification and factory data collection; The bands on or higher than 433 MHz are used for the application with long distance over 1 m. 
The band on 13.56 MHz is used for most RFID cards such as smart cards, memory cards and micro processor cards, which serve our daily life.\cite{dipankar2009rfid}

Protocols on the 13.56 MHz band are ISO 14443 Type A, ISO 14443 Type B, ISO 15693, ISO 18000-3 Mode 1\&2, ISO 18092 NFC, EPC HF CLASS 1 and EPC HF Version 2 etc.
\subsection{Protocols and Products}
ISO 14443 Type A is commonly used for entrance guards, metro cards and consumer stored-value cards. They occupy a high market ratio.

The part of corresponding products are follows:
\begin{itemize}
\item MIFARE Ultralight: also called U10, 64 B memory;
\item MIFARE Ultralight C: also called U20, 192 B memory;
\item MIFARE Classic S50: 1 kB memory;
\item MIFARE Classic S70: 4 kB memory; 
\item MIFARE DESFire family.
\end{itemize}
Among them, there is no cryptography function in MIFARE Ultralight family. They are only memory cards.

The encryption algorithm in MIFARE Classic family is a proprietary algorithm, Crypto-1, created by NXP Semiconductors. The algorithm was attacked by a reverse-engineering and was published in 2008.\cite{Nohl:2008:RCR:1496711.1496724} The initial encrypt key is ``FF FF FF FF FF FF'' or other fix form. A MIFARE Classic card can easily be cracked quickly if there is a known sector key in the card. The most popular crack tool is MIFARE Classic Offline Cracker\cite{MFOC}. Although cards in MIFARE Classic family are not safe, they have been always the mainstream in the RFID market till now for their low costs.

Three well-known public block algorithms DES, 3DES and AES are applied in MIFARE DESFire family to make the cards safer than the ones on previous generation. The initial master key for DES is ``00 00 00 00 00 00 00 00''. There is not any effective attack on MIFARE DESFire family.

%ISO 14443 Type B involves more cryptography functions. Hence it is more suitable to be used for CPU cards such as identity cards, e-passports and bank cards. The part of corresponding products are ST SR176, ST SRIX4K and ATMIL AT88RF020.

%The access distance of ISO 15693 can reach 1 m. The part of corresponding products are ICODE SLI, ICODE SLI-S, ICODE SLI-L, TI Tag-it HF-1 Plus and EM4135.

\section{Attack Model}\label{AttackModel}
We suppose that the attack occurs in a place such as campus or firm using the ``all-in-one'' RFID cards.
And there are two service providers: A and B. They provide different services to a same group of users. A user pays a bill or authenticates his/her identity with a same RFID card.
Because two service are different, they occupy the different storage spaces (also called sectors or applications) on an RFID card.
For the safety of their data, service providers always set access keys on their users' cards. and the keys are strictly confidential.
Meanwhile, for users, all these things are transparent. Only when they view the balance or pass an access control, they can know whether their RFID cards are available.

\begin{figure}[ht]
    \centering
    \includegraphics[width=0.5\textwidth]{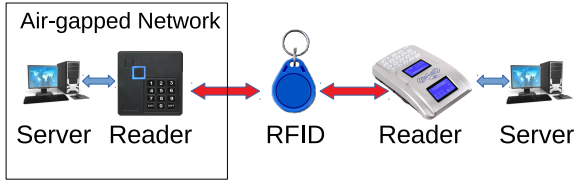} 
    \caption{Schematic Diagram of Attack Model}
    \label{SchematicDiagram}
\end{figure}

However, when two service providers are at different security levels, the design of this type of multi-purpose cards for convenience has a great security risk.

\begin{figure}[ht]
    \centering
    \begin{subfigure}[b]{0.5\textwidth}
    \centering
        \includegraphics[width=0.6\textwidth]{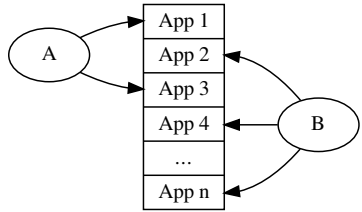}
        \caption{Safe Access by Two Service Providers}
        \label{safeaccess}
    \end{subfigure}
    
    \begin{subfigure}[b]{0.5\textwidth}
    \centering
        \includegraphics[width=0.6\textwidth]{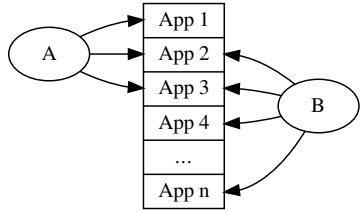}
        \caption{Unsafe Access by Two Service Providers}
        \label{unsafeaccess}
    \end{subfigure}
\caption{Two Access Form by Two Service Providers}
\end{figure}

Shown in Figures \ref{SchematicDiagram} and \ref{safeaccess}, we suppose that the service of A is in a high security level. Its card readers and server are settled in an air-gapped network. The server connects with other computers in order to provide effective data information to the users in time, such as attendance information on an entrance guard system etc.
In comparison, the service of B is in a lower security level or a public level. For ease of management and operation, the readers and server connect with the less security network or Internet. For example, the service can be an RFID card toll system in a restaurant or a canteen.

Shown in Figure \ref{unsafeaccess}, if A colludes with B, they can communicate each other covertly by accessing RFID cards. Then the sensitive data in a high security network would be leaked to a lower security network or Internet. Generally, the storage spaces can be accessed with the default key if the access keys to the storage spaces have not been changed. Therefore, A and B can exchange information by accessing these spaces with the default key. They can also change the keys to these spaces to avoid detection and keep their messages secret. Even if all the keys to other spaces have been changed by others, A and B can also communicate covertly by using the spaces assigned to them.

Even if A and B do not collude with each other, if their keys are obtained with other ways, the attacker can also transmit data covertly in the normal business procedure by reprogramming the execution codes of the readers.

Now a prototype of such attacks is proposed.

\subsection{Frame Structure}
The capacity of the storage spaces accessed by A and B to transmit message is very small. Therefore, large-sized data should be divided into groups before transmission.

\begin{figure}[ht]
\includegraphics[width=0.5\textwidth]{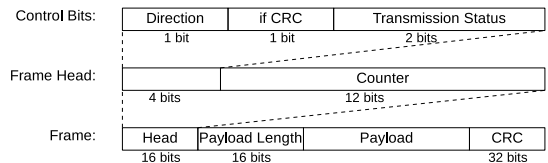} 
\caption{Frame Structure}
\label{FrameStructure}
\end{figure}

For an efficient packet transmission, the packet frame is designed. The structure of a packet frame is shown in Figure \ref{FrameStructure}.

Bit 1 in the frame represents the transmission direction: `0' represents from A to B; `1' represents from B to A. Bit 2 means that whether CRC is used in frame tail. Bits 3$\sim$4 represent transmission status listed in Table \ref{TransmissionStatus}.  Bits 5$\sim$16 are formed a frame counter. Bits 17$\sim$32 are used to record the payload length. There is a 32-bit CRC in the frame tail if Bit 2 of the frame head is `1'.

\begin{table}[ht]
\caption{Transmission Status}
\label{TransmissionStatus}
\center
\begin{tabular}{c|c|c}
\hline 
Bits & Transmission Status & Payload\\
\hline 
00 & Normal Transmission & Yes\\
01 & Succeed to Transmit & No\\
10 & Ask for Transmit Again & No\\
11 & Query for Transmission Status & No\\
\hline 
\end{tabular}
\end{table}

Therefore, the frame is variable-length. Supposing that the capacity of the storage spaces accessed by A and B is $s$ bytes, beside the 2-byte frame head and the 2-byte payload length, there are almost $s-4$ bytes for the payload of a frame.

Once a frame is found on an RFID card by the receiver, the whole frame would be read from the card according to the length recorded in the frame head. Then, the CRC will be verified if Bit 2 is `1'. If there is no error, the payload is stored in the computer with the order of the frame counter. And Bit 4 of the frame head in the card will be set as `1'; If there is something wrong, the payload will be abandoned. And Bit 3 of the frame head in the card will be set as `1'. Bit 2 is set as `0', that is, no CRC is needed.

To improve transmission efficiency, if there are some data needed to be sent back, the receiver can use the rest of storage space in length of $s-2$ bytes to send the data in frames.
In the same way, if the data package sent by receiver is lost, such as the card user asked for a leave for a while, the sender have not received the transmission status of some frame. Then the sender can query for the transmission status actively. At the moment, Bits 2$\sim$4 are set as `011'. And the frame counter is set as the frame number for which will be queried. The length of payload is set as zero. The sender can also send more frame in the rest of storage space.

In short, frames can be cascaded as long as there is enough space in the card.

\subsection{Channel Rate}
We suppose that there are $c$ users who use the services of A and B frequently. And in average, one user use the two services alternately for $f$ times per day. Then, in a whole day, the capacity of the covert communication between A and B is:
\[2\times c\times f\times s ~\text{bytes}\]

Getting rid of the frame heads and tails, the actual maximum capacity for the payloads is: 
\[2\times c\times f\times (s-6) ~\text{bytes}\]

\section{Results and Discussion}\label{ResultandDiscussion}
To test the effectiveness of this attack, we used mainstream RFID cards and card readers on the market. And we developed two RFID cards service simulation systems. The service systems are used to simulate the attack scenario described in the previous section.
\subsection{Experimental Setting}
Hardware:

\begin{itemize}
\item RFID cards: MIFARE Classic S50 $\times$ 10;
\item Card readers: PN532 $\times$ 2;
\item 2 USB-TTL adapters.
\end{itemize}

Software:
\begin{itemize}
\item Operating System: Ubuntu 18.04;
\item Driver: libnfc;
\item Program Language: GNU C;
\item Database: SQLite.
\end{itemize}

\begin{figure}[ht]
\includegraphics[width=0.5\textwidth]{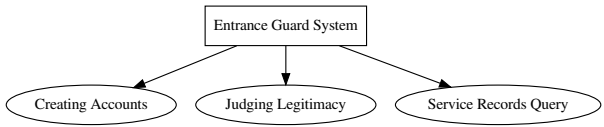} 
\caption{Functions in Entrance Guard System}
\label{EntranceGuard}
\end{figure}

\begin{figure}[ht]
\includegraphics[width=0.5\textwidth]{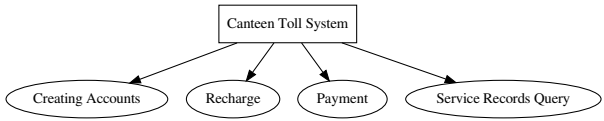} 
\caption{Functions in Canteen Toll System}
\label{CanteenToll}
\end{figure}

The two systems: entrance guard system and canteen toll system, were developed with the software configuration above. Shown in Figure \ref{EntranceGuard} and Figure \ref{CanteenToll}, functions of the entrance guard system involve creating accounts, judging user's legitimacy and service records query; and functions of the canteen toll system involve creating accounts, recharge, payment and service records query.

Now, we use 10 RFID cards to create accounts on both systems. That is, these 10 cards can be used on both systems at the same time. Then we simulated the attackers, tampered with the function to judge user's legitimacy in the access control system and the payment function in the canteen toll system. Malicious codes are injected in so that the card reader could read and write to other storage spaces while working normally.

There are 16 sectors in a MIFARE Classic S50 card. One sector is made up of 4 blocks. Except the first block in the first sector and the last block in every sector, the rest blocks can be used to read and write data by using the keys to their sectors. Because only the first block in the first sector is needed to read by the entrance guard system to find the tag ID. Hence no block space is needed to store data in the system; On the other hand, the first block and the third block in the third sector are used by the canteen toll system to store a username and the balance respectively. Therefore, there are 45 blocks that can be exploited for covert communications.

We suppose that on weekdays these 10 users go through the entrance guard system and start their work in the morning, and have their lunch after payment on the canteen toll system, then continue their work in the afternoon passing the entrance guard system, finally have their dinner in the canteen after their work. It's a very common scene in big cities. Thus, one user uses the both services alternately for 2 times in a whole day.

\subsection{Results}
According to the experimental settings, we know that:
\[s = 45\times 16 = 720~\text{bytes}\]
\[c = 10\]
\[f = 2\]
Therefore, the capacity of the covert communication between both systems in a whole day is $2\times 720 \times 10 \times 2 = 28800$ bytes. This number is small, but it is enough to transmit secret information such as passwords, private keys and credit card numbers etc.

The daily capacities of the covert communication via different type of RFID cards are listed in Table \ref{TransmissionCapacityofDifferentType} when 10 users use the both services alternately for 2 times.

\begin{table}[ht]
\caption{Daily Capacities of Different Type of RFID Cards}
\label{TransmissionCapacityofDifferentType}
\center
\begin{tabular}{c|c|c}
\hline
Card Type& Total Cap. & Transmission Cap.\\
\hline
S50 & 1,024 B & 28,800 B\\
S70 & 4,096 B & 120,960 B\\
D21 & 2,048 B & 71,680 B\\
D41 & 4,096 B & 145,920 B\\
D81 & 8,192 B & 294,400 B\\
\hline
\end{tabular}
\end{table}

\subsection{Discussion}
Reasons for the existence of this type of covert channels are follows:

Firstly, the access key is the only evidence to verify the validity of a card reader to the card.
The idea comes from cryptography, but is not very suitable to access control of RFID.

Secondly, the design of RFID cards does not match with the security requirement in the ``all-in-one'' mode. In the ``all-in-one'' mode, it is usual that several service providers access a same card. And the owner of a reader is not fixed in the long term. When some service providers quit, they can sell part or all of their readers to other service providers.

Finally, collusion between different service providers is considered unlikely. In the view of cryptography, once a service provider shares its access keys to others, the security of its data is not guaranteed. Therefore, no service provider would share its access keys actively. However, there are a lot of situations under which the access keys of a service provider is shared. Some general situations are follows:

\begin{itemize}
\item The service provider is compromised by a hacker;
\item The service provider mistakenly believed the assurance of another;
\item The benefits of sharing the keys far outweigh the possible losses.
\end{itemize}

Generally, the evidence to verify the validity of a card reader to a card should be something that can not be copied easily.

\section{Proposed Protocol}\label{ProposedProtocol}
To avoid the covert communication between service providers via RFID cards, a trusted third party, an RFID card administrator, is introduced. Therefore, four parties take part in the new protocol. They are:

\begin{itemize}
    \item RFID Card (Tag);
    \item RFID Card Reader (Reader);
    \item Service Provider (SP);
    \item Card Administrator (CA).
\end{itemize}   

The proposed protocol assures technically that the four parties can achieve the following objectives.
\begin{itemize}
\item An SP can only access the storage space on an RFID card according to the settings of the CA.
\item When a reader is needed to added or changed by an SP, the reader must be registered with the CA to get an encrypted command.
\item Different SPs can not use the same card reader serial number.
\item Different SPs can not access the same storage space on RFID cards. 
\end{itemize}

\subsection{Data Storage}

\subsubsection{Tag}
The data storage form on a tag is listed in Table \ref{DataStorageFormsofTags}, where TID is the serial number of tag. CA Key is the common key with CA. Only CA can write it into the card. Neither card holders nor SPs can access it. Master Key is the key to access the card. Both the CA and SPs know the master keys to their users' cards. 

\begin{table}[ht]
\caption{Data Storage Form on Tags}
\label{DataStorageFormsofTags}
\center
\begin{tabular}{|c|}
\hline
TID\\
\hline
CA Key\\
\hline
Master Key\\
\hline
\hline
App1 Version\\
\hline
App1 White List\\
\hline
App1 App Key\\
\hline
Data of App1\\
\hline
\hline
App2 Version\\
\hline
App2 White List\\
\hline
App2 App Key\\
\hline
Data of App2\\
\hline
\hline
$\cdots$\\
\hline
\end{tabular}
\end{table}

For an app in the card, App White List is a list of serial numbers and frequentness values of card readers. A card reader can access an application in the card with a correct app key only if the serial numbers of the reader is in the white list of the application. Otherwise, the tag will require for an encrypted command $E_{CAK}(RID, ID_{App}, version)$. If the card reader can not present a valid encrypted command, it can not access the application even if the app key is correct. 
App Version is a value to remark the current version of encrypted commands in the white list. When a new valid encrypted command comes, there are three possible statues by comparing the version value in the new command($CV$) and the App Version value($AV$):

\begin{enumerate}
\item If $CV = AV$, the values of $RID$ and $ID_{App}$ will be inserted in the white list;
\item If $CV < AV$, it means that the new-coming encrypted command is out-of-date.
\item If $CV > AV$, it means that the CA has updated the version. And all items in the white list are out-of-date. Then the white list will be cleared, and the values of $RID$ and $ID_{App}$ will be inserted in the white list.
\end{enumerate}

Due to the limited storage space in an RFID card, an app white list should be designed as a fixed-length list. A item in the list includes the 28-bit RID and 4-bit frequentness. The initial frequentness value is 8. When the reader with RID access the app, the frequentness is added by 1. Meanwhile, the tag chooses randomly one other frequentness values that is decreased by 1. If the frequentness value is already 15, it will not be added and none of other frequentness values is decreased.

When the white list is full, the RID in list with the minimal frequentness value will be replaced with a new coming RID. The frequentness value of the new RID is also the the minimal value; And when the white list is cleared, all frequentness values are reset as the initial value, that is 8.

\subsubsection{Reader}
The data storage form on a card reader is listed in Table \ref{DataStorageFormsofReaders}. RID is the serial number of reader. The list of app keys and encrypted commands $E_{CAK}(RID, ID_{App}, version)$ to the applications are stored in a card reader, where $version$ is an integer that will be increased when the SP of the readers is changed.

\begin{table}[ht]
\caption{Data Storage Form on Readers}
\label{DataStorageFormsofReaders}
\center
\begin{tabular}{|c|}
\hline
RID\\
\hline
\hline
$ID_{APP1}$\\
\hline
App1 App Key\\
\hline
$E_{CAK}(RID, ID_{App1}, version)$\\
\hline
\hline
$ID_{APP2}$\\
\hline
App2 App Key\\
\hline
$E_{CAK}(RID, ID_{App2}, version)$\\
\hline
\hline
$\cdots$\\
\hline
\end{tabular}
\end{table}

\subsubsection{SP}

The data stored in SPs' servers are in the form of a database table. The database table is used to record the information of all tags with two fields: serial number(TID) and master key.

\begin{table}
\caption{Data Storage Form on SP}
\begin{center}
\begin{tabular}{|c|c|}
\hline
TID & Master Key\\
\hline
$\cdots$ & $\cdots$\\
\hline
\end{tabular}
\end{center}
\end{table}

\subsubsection{CA}
The data in the CA's server are threefold: CA Key and two database tables. One database table is used to record the information of all tags as same as the database table in SPs' servers in the form; The other database table is used to record the settings of app distribution to the readers with four fields: serial number of SP($ID_{SP}$), serial number of reader(RID), serial number of app($ID_{APP}$) and version.

\begin{table}
\caption{Data Storage Form on CA}
\begin{center}
\begin{tabular}{|c|}
\hline
CA Key\\
\hline
\end{tabular}

\begin{tabular}{|c|c|}
\hline
TID & Master Key\\
\hline
$\cdots$ & $\cdots$\\
\hline
\end{tabular}

\begin{tabular}{|c|c|c|c|}
\hline
$ID_{SP}$ & RID & $ID_{APP}$ & version\\
\hline
$\cdots$ & $\cdots$ & $\cdots$ & $\cdots$\\
\hline
\end{tabular}
\end{center}
\end{table}

\subsection{Initialization}
The procedure of initialization is divided into two aspects: issuing cards to users and authenticating the card readers.

When a card is issued to a user, the CA would write the common key(CA Key) into the RFID card. Then the card can decrypt the encrypted command to verify the reader. 

Meanwhile, every reader of all SPs must be registered with the CA. The CA obtains the serial number of a reader(RID) and determines the application number that will be accessed by the reader($ID_{APP}$). Then an encrypted command is created by encrypting the content of RID, $ID_{APP}$ and $version$ with the CA Key. Finally, the CA writes the encrypted command into the reader.

\subsection{Authentication Procedure}
\begin{figure*}[ht]
\center
\includegraphics[width=0.8\textwidth]{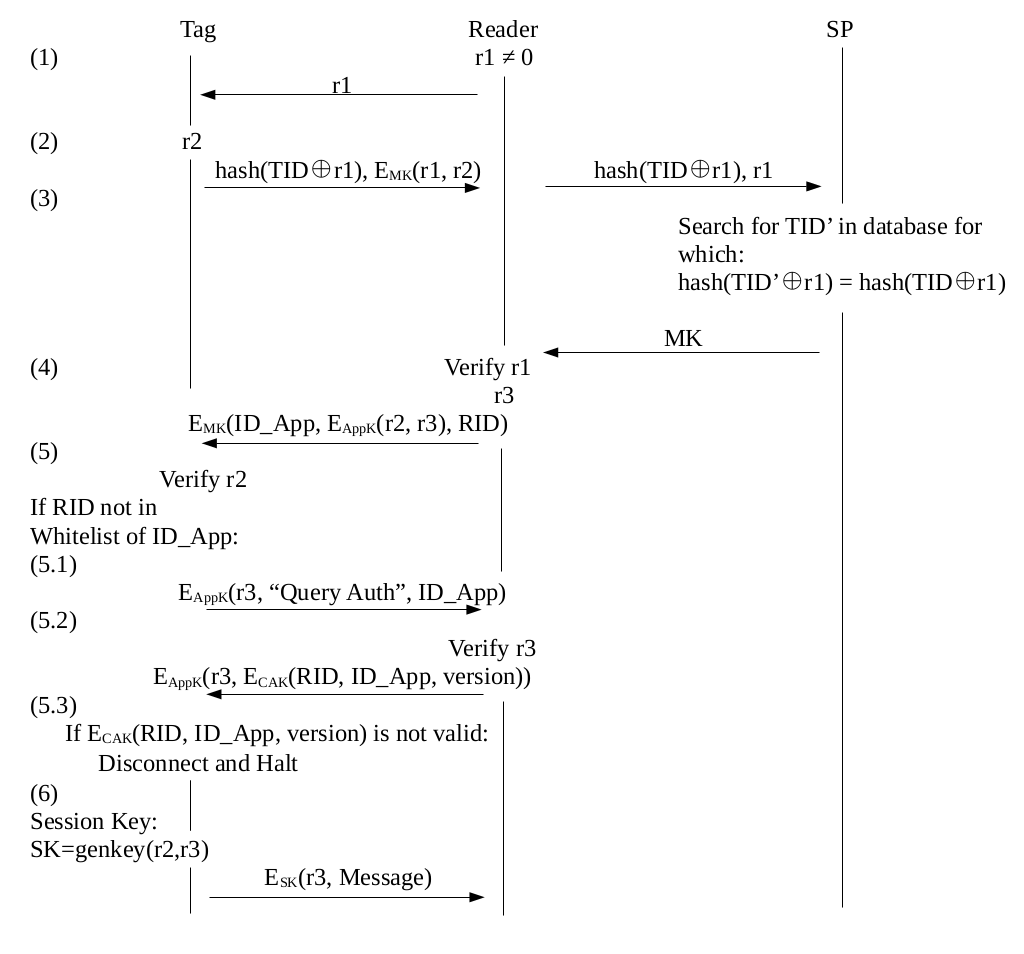} 
\caption{New Authentication Protocol}
\label{NewAuthenticationProtocol}
\end{figure*}

The authentication procedure is described in Figure \ref{NewAuthenticationProtocol}.
This protocol is different from the previous authentication protocols. The RFID card not only checks whether the reader has the master key and app key of the RFID card, but also checks whether the reader has a privilege to access related applications.

\textbf{Step 1:} Reader creates a nonzero random number $r_1$, and transmit it to Tag.% and serial number of Reader(RID)

\textbf{Step 2:} Tag receives $r_1$, and creates another random number $r_2$. Then Tag encrypts them as $E_{MK}(r_1, r_2)$ with the master key(MK). And Tag calculate the hash value of exclusive or(XOR) of its serial number(TID) and $r_1$. Then Tag sends $hash(TID \oplus r_1)$ and $E_{MK}(r_1, r_2)$ to Reader.

\textbf{Step 3:} Reader forwards the message $hash(TID \oplus r_1)$ and $r_1$ to SP. SP looks up its database table, query the TID by calculating the value of $hash(TID \oplus r_1)$ for all the TIDs in the table. If the TID is found, SP sends the MK of the TID back to Reader.

\textbf{Step 4:} Reader decrypts $E_{MK}(r_1, r_2)$ with the MK, and checks the value of $r_1$. If it is equal to the initial value, it means that Tag is valid. Then Reader creates a random number $r_3$, and encrypts $r_2$ and $r_3$ as $E_{AppK}(r_2, r_3)$ with the application key(AppK), and encrypts the serial number of app($ID_{APP}$) that is going to be accessed, $E_{AppK}(r_2, r_3)$ and its serial number(RID) with the MK as $E_{MK}(ID_{APP}, E_{AppK}(r_2, r_3), RID)$. Then Reader sends it to Tag.

\textbf{Step 5: }
\begin{description}
\item[Step 5.1] Tag decrypts $E_{MK}(ID_{APP}, E_{AppK}(r_2, r_3), RID)$ with MK, and decrypts $E_{AppK}(r_2, r_3)$ with the application key according to $ID_{APP}$. Then Tag verifies $r_2$. If $r_2$ is correct, Tag judges whether RID is in the white list of App $ID_{APP}$. If not, Tag encrypts $r_3$, the words ``Query Auth'' and $ID_{APP}$ as $E_{AppK}(r3, ``\text{Query Auth}", ID_{APP})$. Then Tag sends it to Reader; 
\item[Step 5.2] Reader decrypts it, and verifies $r_3$. If $r_3$ is correct, Reader queries an encrypted command in its memory according $ID_{APP}$. If the encrypted command is found, Reader encrypts $r_3$ and the encrypted command as $E_{AppK}(r3, E_{CAK}(RID, ID_{APP}, version))$ with the application key, and sends it to Tag; 
\item[Step 5.3] Tag decrypts it, and decrypts $E_{CAK}(RID, ID_{APP}, version)$ with the CA key(CAK). Then Tag judges whether the encrypted command is valid. The command is valid if and only if ``the RID and $ID_{APP}$ is matched with the session context, and the value of $version$ is not less than the value of the App Version''. If not, Tag disconnects the link with Reader and halts at once. Otherwise, if the value of $version$ is greater than the value of the App Version, the white list will be cleared. Finally, the value of RID is inserted in the white list. And the value of the App Version is updated as the value of $version$.
\end{description}

\textbf{Step 6:} Tag and Reader calculate the session key $SK$ by the values of $r2$ and $r3$. Then they encrypt the messages with the session key to start further normal business process.

\subsection{Formal Security Verification Using ProVerif}
We use ProVerif\cite{ProVerif}, an automatic cryptographic protocol verifier in the formal model, to verify the proposed protocol. Firstly, a public channel and a basic type of variables are defined. Then, the cryptography functions of the proposed protocol is modeled, and the secret keys, events and authentication queries are defined. Thirdly, the processes of the tag and the reader are modeled respectively. Finally, the whole process of the proposed protocol is modeled.

The ProVerif code of the proposed protocol is listed in the Appendix, and the simulation result with ProVerif v2.00 is as follows.

\lstset{ 
	basicstyle=\footnotesize,
	breaklines=true,
}
\begin{lstlisting}
RESULT inj-event(Tagend(x)) ==> inj-event(Tagbegin(x)) is true.
RESULT inj-event(Readerend(x_30)) ==> inj-event(Readerbegin(x_30)) is true.
RESULT not attacker(Tname[]) is true.
RESULT not attacker(Rname[]) is true.
\end{lstlisting}

The result demonstrates that the proposed protocol satisfies the security requirement of session key and achieves mutual authentication successfully.

\subsection{Evaluations}
The proposed protocol solves the problem that different SPs communicate with each other covertly via RFID cards. However, some security matters still should be taken care.

For example, an SP tampered with the sequence number of card reader, and obtained corresponding encrypted command. Hence any tag regards the reader as another valid one.
Then, the SP puts two readers into one reader shell. Once the normal work is done, the first reader shuts down, and the second one starts to work immediately. Therefore, the whole security mechanism is compromised.

In response, the continuous access from different readers in a few seconds should be forbidden by changing the circuit of RFID. If a card reader always requires the users to swipe their RFID cards twice, the phenomenon will make people's awareness. Thus, the covertness of the communication can not be guaranteed. 

\section{Conclusions}\label{Conclusions}
In the paper, a bidirectional covert channel via multipurpose RFID cards between service providers was established. 
An attack prototype of information leakage was implemented between two simulation systems by using the RFID cards in MIFARE Classic and DESFire families. 
The daily channel capacity to leak data was obtained by calculations and experiments. Furthermore, the reasons for existence of such covert channels were discussed. 
To eliminate this type of covert channels, a new authentication protocol between RFID cards and card readers were proposed.
Our results show that the new authentication protocol can effectively avoid covert communication between service providers via RFID cards while keeping user convenience.

\section*{Acknowledgments}
This work was supported in part by the Natural Science Foundation of China under Grant U1636201 and 61572452, and by Anhui Initiative in Quantum Information Technologies under Grant AHY150400.

\section*{Web Resources}

Research Site: \url{http://home.ustc.edu.cn/~zhou7905/RFIDCC}

Demo Video: \url{https://www.youtube.com/watch?v=b1iPeC4O56k}

\Urlmuskip=0mu plus 1mu\relax
\bibliographystyle{plain}
\bibliography{ref.bib}

\section*{Appendix}
\subsection{ProVerif Code}
\scriptsize
\lstset{ 
	basicstyle=\footnotesize,
	breaklines=true,
}
\begin{lstlisting}
(* Private channel *)
free ch1:channel [private].
free ch2:channel [private].

(* Public channel *)
free c:channel.

(* Symmetric key encryption *)
type key.
fun senc(bitstring, key): bitstring.
reduc forall m: bitstring, k: key; sdec(senc(m,k),k) = m.

type role.
free tag, reader: role.

(* Symmetric key generation function *)
fun genkey(bitstring, bitstring): key.

(* Hash function *)
fun hash(bitstring): bitstring.

(* Xor function *)
fun xor(bitstring, bitstring): bitstring.

(* Authentication queries *)
event Tagbegin(role).
event Tagend(role).
event Readerbegin(role).
event Readerend(role).
query x: role; inj-event(Tagend(x)) ==> inj-event(Tagbegin(x)).
query x: role; inj-event(Readerend(x)) ==> inj-event(Readerbegin(x)).

free Tname, Rname:bitstring [private].
query attacker(Tname);
	attacker(Rname).

(* CA *)
let processCA() =
	new CAK: key;
	new MK: key;
	new AppK: key;
	in(ch1, TID: bitstring);
	out(ch1, (CAK, MK, AppK));
	in(ch2, RID: bitstring);
	new ID_APP: bitstring;
	new version: bitstring;
	out(ch2, senc((RID, ID_APP, version), CAK));
	out(ch2, (ID_APP, MK, AppK)).

(* Tag *)
let processT() =
	new TID: bitstring;
	out(ch1, TID);
	in(ch1, (CAK: key, MK: key, AppK: key));
	event Tagbegin(tag);
	in(c, r1: bitstring);
	new r2: bitstring;
	out(c, (hash(xor(TID, r1)), senc((r1, r2), MK)));
	in(c, EMK:bitstring);
	let (ID_APP: bitstring, EAppK: bitstring, RID: bitstring) = sdec(EMK, MK) in
	let (=r2, r3: bitstring) = sdec(EAppK, AppK) in
	out(c, senc((r3, ID_APP), AppK));
	in(c, EAppK2: bitstring);
	let (=r3, ECAK: bitstring) = sdec(EAppK2, AppK) in
	let (=RID, =ID_APP, version: bitstring) = sdec(ECAK, CAK) in
		event Readerend(reader);
	out(c, senc(Tname, genkey(r2, r3))).

(* Reader *)
let processR() =
	new RID: bitstring;
	out(ch2, RID);
	in(ch2, ECAK: bitstring);
	in(ch2, (ID_APP: bitstring, MK: key, AppK: key));
	event Readerbegin(reader);
	new r1: bitstring;
	out(c, r1);
	in(c, (HTID: bitstring, EMK: bitstring));
	let (=r1, r2: bitstring) = sdec(EMK, MK) in
	new r3: bitstring;
	out(c, senc((ID_APP, senc((r2, r3), AppK), RID), MK));
	in(c, EAppK: bitstring);
	let (=r3, =ID_APP) = sdec(EMK, MK) in
		event Tagend(tag);
	out(c, senc((r3, ECAK), AppK));
	out(c, senc(Rname, genkey(r2, r3))).

process
	(  (!processCA()) |(!processT()) | (!processR()) )
\end{lstlisting}

\end{document}